\documentclass[12pt]{article}
\usepackage{graphicx,subfigure,epsfig}
\usepackage{color}

\begin{document}

\title{Interface-dominated Growth of a Metastable Novel Alloy Phase}

\author{Subhendu Sarkar, Alokmay Datta, Purushottam Chakraborty\\
\textit{Surface Physics Division,}\\\textit{Saha Institute of
Nuclear Physics,}
\\\textit{1/AF Bidhannagar, Kolkata 700 064, India.}\\ and\\Biswarup Satpati\\
Institute of Physics, \\\textit{Sachivalaya Marg, Bhubaneswar 751
005, India.}}

\maketitle


\begin{abstract}
A new \textit{D0$_{23}$} metastable phase of Cu$_3$Au is found to
grow at the interfaces of Au/Cu multilayers deposited by magnetron
sputtering. The extent of formation of this novel alloy phase
depends upon an optimal range of interfacial width primarily
governed by the deposition wattage of the dc-magnetron used. Such
interfacially confined growth is utilized to grow a $\sim$ 300 nm
thick Au/Cu multilayer with thickness of each layer nearly equal
to the optimal interfacial width which was obtained from secondary
ion mass spectrometry (SIMS) data. This growth technique is
observed to enhance the formation of the novel alloy phase to a
considerable extent. SIMS depth profile also indicates that the
mass fragment corresponding to Cu$_3$Au occupies the whole film
while x-ray diffraction (XRD) shows almost all the strong peaks
belonging to the \textit{D0$_{23}$} structure. High resolution
cross-sectional transmission electron microscopy (HR-XTEM) shows
the near perfect growth of the individual layers and also the
lattice image of the alloy phase in the interfacial region. Vacuum
annealing of the alloy film and XRD studies indicate stabilization
of the \textit{D0$_{23}$} phase at $\sim$ 150$^{\circ}$C. The role
of interfacial confinement, the interplay between interfacial
strain and free energy and the hyperthermal species generated
during the sputtering process are discussed.

\end{abstract}

\newpage

\section{Introduction}
In metallic multilayer systems the large free energies available
at the interfaces can help to form interfacial alloys when these
free energies are comparable to or larger than the enthalpy of
formation of the probable alloy phases~\cite{fournee, davies}. In
many such systems the alloy formed is of a non-bulk, metastable
nature~\cite{kong, zhang, abadias}. For multilayers of immiscible
metals there exists a positive mixing enthalpy that in effect acts
as an energy barrier to mixing~\cite{yavari, he}. On the other
hand for miscible metallic systems the formation enthalpy for
different metastable crystalline or ordered phases are mostly
negative and very close to each other~\cite{zun}. Interfacial
alloy formation in multilayers have been reported to take place
via solid state reaction~\cite{chen, schwarz}, mechanical
alloying~\cite{mori, dorofeev, ma}, ion beam irradiation
~\cite{koike, hung} etc. Especially in case of the immiscible
systems, alloying is achieved via ion beam irradiation technique,
which can supply adequate energy to overcome the energy barrier.
Almost all of the metastable alloys reported for immiscible
systems are amorphous in nature, the basic idea being the
persistence of a random arrangement of atoms formed by forced
mixing and subsequent rapid quenching of the supplied
energy~\cite{hafner}. The probability of obtaining ordered alloy
phases increases as we go over to the miscible systems. Local
density approximation (LDA) calculations for Au/Cu systems have
shown that a number of ordered metastable phases have minima of
formation enthalpy close to \textit{L1$_{2}$}, the ordered stable
phase of the gold-copper alloy Cu$_3$Au in bulk~\cite{zun}, and
indeed the first indication of a metastable phase of Cu$_3$Au was
observed in a thermal study of the bulk alloy~\cite{cohen}.
Metastable alloys could be easier to form at Au/Cu interfaces, and
sputter deposition techniques, because of their high quench rate,
can be more effective in achieving such a situation. Furthermore,
hyperthermal atoms formed during the sputter deposition process
can facilitate to modify or create new phases present at the
interfaces~\cite{hanley,zhou}.

In view of the above facts, we had previously studied (Au/Cu)$_n$,
(n = 2,10) multilayers deposited on glass and Si (001) at ambient
temperature by dc-magnetron sputtering~\cite{mycu3au}. Figure~1
summarizes the results obtained from that study. The nominal
thickness of each layer was $\sim$ 50 nm (Figure~1a). Secondary ion
mass spectrometry (SIMS) (Figure~1b), and high-resolution
cross-sectional transmission electron microscopy (HR-XTEM) studies
showed that the multilayer periodicity and the crystallinity of the
individual layers were excellent. The double peaks in the Au signal
at the multilayer interfaces observed in the SIMS profile are due to
the presence of oxygen trapped preferentially in the Cu layers. This
observation is attributed to be due to the `matrix
effect'~\cite{myoxy}, a common phenomenon in SIMS. The crystallinity
of the individual layers were also brought out by the fact that the
layers were highly oriented in the $\langle$111$\rangle$ direction
as obtained from the x-ray diffraction (XRD) results (Figure~1c).
But the most interesting result of the study~\cite{mycu3au} was that
a \textit{D0$_{23}$} phase of Cu$_3$Au alloy (Figure~1d), a
metastable ordered phase predicted from LDA calculations~\cite{zun}
formed only at the multilayer interfaces. The interfacial location
of the alloy was confirmed from SIMS results (inset of Figure~1b)
where the Cu$_2$Au fragment of Cu$_3$Au alloy was seen to be
confined only across the interfacial region. This was the first
experimental finding of Cu$_3$Au in the \textit{D0$_{23}$} phase
rather than its bulk \textit{L1$_{2}$} form (Figure~1d). It was also
confirmed that formation of this novel phase was not a direct
outcome of the deposition process but primarily due to confinement
at the multilayer interfaces~\cite{mycu3au}. The major hurdle
towards studying this phase in greater detail was practically due to
the insufficiency in the amount of this alloy phase. Hence, a proper
structural confirmation of the \textit{D0$_{23}$} phase could not be
completed in that case. We therefore planned to overcome this
difficulty through enhancement of the formation of this metastable
alloy utilizing its growth under interfacial confinement.

In this communication we present a novel, simple but effective
method to obtain the \textit{D0$_{23}$} phase of Cu$_3$Au in
sizable amounts. As Au-Cu pair is one of the commonest pairs of
miscible metals and as the method is based on optimizing the
interfacial growth of the alloy, we believe that this method can
be applicable to metastable interfacial alloys of miscible metals
in general. We have used this method to grow Cu$_3$Au alloy in the
\textit{D0$_{23}$} phase in considerable amounts in a $\sim$ 300
nm thick Au/Cu multilayer and have been able to confirm its
structure from the number of new peaks it provided in x-ray
diffraction. Upon vacuum annealing, this alloy phase stabilizes at
around 150$^\circ$C, consistent with the higher enthalpy minimum
of \textit{D0$_{23}$} compared to \textit{L1$_{2}$}, as found from
LDA calculations~\cite{zun}.

\section{Experimental}
As a first and necessary step towards achieving the
interface-dominated growth, we needed to estimate the optimal
conditions favoring the growth of this metastable alloy phase.
Towards this end, we deposited (Cu/Au)$_2$ multilayers on glass at
ambient temperature by dc-magnetron sputtering in a Pfeiffer
PLS500 sputter-coating unit. The multilayers were grown on the
substrates (kept at room temperature) by sequential deposition of
Au and Cu. During deposition the substrates attached to the sample
holder were rotated to ensure uniform film growth. The geometry of
the deposition chamber was such that the depositing atoms arrive
at an angle normal to the substrate. The distance between the
target and the substrate was 11.0 cm. The base pressure was
3.4$\times$10$^{-6}$ mbar and prior Ar (purity $\sim$ 99.99\%,
BOC) flushing minimized impurities during deposition. The working
pressure was in the range 3.5 -- 5.0$\times$10$^{-3}$ mbar. The
nominal thickness for each layer was $\sim$ 50 nm. The substrates
were cleaned thoroughly with trichloroethylene and methanol. The
Ar gas flow rate was 10 sccm at the time of deposition. These
parameters were identical to those employed in Ref.
\cite{mycu3au}. Three sets of samples were deposited at three
different wattages viz., 13, 25 and 50 watts. SIMS studies were
carried out using a QMS-based SIMS instrument (HIDEN Analytical
Ltd, UK) with a high performance triple quadrupole filter and a
45$^\circ$ electrostatic sector-field energy
analyser~\cite{mymcs}. Bombardment was done with cesium ions (1
keV, 50 nA) and secondary negative ions were detected during the
measurement. X-ray diffraction (XRD) was done using a Philips
X-ray diffractometer with a fixed anode-type Cu K$_\alpha$ source
of wavelength 1.54 \AA. The power of the source was 800 W. The
diffraction angle 2$\theta$ was varied from 10$^\circ$ to
100$^\circ$. The minimum step size (2$\theta$) was 10 mdeg. For
cross-sectional TEM (XTEM) studies, samples were prepared using a
wire saw (SBT 850), a lapping and polishing machine (SBT 910), a
dimple grinder (Gatan 656) and a precision ion polishing system
(Gatan 691). The samples were lapped and dimpled up to 120 and 40
$\mu$m, respectively. Ion polishing was done with Ar ions at 3 keV
up to electron transparency. The TEM measurements were carried out
in a JEOL JEM-2010 transmission electron microscope operating at
200 keV.

\section{Results and Discussion}

\subsection{Optimization of the growth parameters}

XRD spectra for samples deposited at the different wattages are
shown in Figure~2a. The position of the (111) peak of the Cu$_3$Au
alloy in the \textit{D0$_{23}$} phase is indicated by an arrow. The
peak to background ratio of the alloy formed at different wattages
is tabulated in table~\ref{table:watt}. This ratio has been
normalized with respect to that for the sample deposited at 50
watts. It is clear from the table that the alloy is better formed at
25 watts since its (111) peak intensity is the strongest at this
wattage. Alloy formation is hindered for both higher and lower
wattage conditions. SIMS results (Figure~2b), on the other hand,
indicate that the interfacial width increases with the deposition
wattage. This is quantitatively shown in table~\ref{table:watt}. The
interfacial widths have been measured from the full width at half
maximum (FWHM) of the individual alloy peaks located only at the
interfaces. Higher deposition wattage indicates deposition at a
higher energy, resulting in the broadening of the interface between
two deposited layers. However, from this table and Figure~2a it is
clear that the \textit{D0$_{23}$} phase is well formed only when the
interfacial width is $\sim$ 7.0 nm.

It is expected that the formation of this interfacial alloy phase
will be enhanced as the number of interfacial atoms as a fraction
of the total number of atoms is increased~\cite{chen, chen1}.
However, for our case this does not seem to be the deciding
factor, since the above fraction, given by the interface width as
a fraction of total thickness, monotonically increases with
increase in deposition wattage. The deciding factor here is rather
a sharply defined degree of confinement for which the particular
alloy is best formed.

\subsection{Growth and characterizations of the interface-dominated alloy }

In view of the above findings, we planned to prepare a multilayer
having identical interfacial confinement conditions but with
individual layer thickness reduced to about the thickness of the
optimum interface. Figure~3 summarizes our interface-dominated
growth process and the results of SIMS and XRD studies on the grown
film. We deposited (Cu/Au)$_{20}$ multilayers on glass by
dc-magnetron sputtering with identical growth conditions as that of
our previous set of samples of two and ten bilayers as already
discussed in the experimental section. The well-defined periodicity
of the multilayer was confirmed from SIMS and XTEM studies. The
individual layer thickness for the present multilayer was about 8.0
nm (Figure~3a). We believed that such low individual thickness would
make up an interface-dominated multilayer stack thereby maintaining
the proper strain so as to form the tetragonal alloy almost
throughout the entire multilayer. SIMS depth profile (Figure~3b,
top) of the Cu signal of the multilayer indicates that the stack has
a well-defined periodicity with all the 20 layers distinctly
separated. More important, however, is the fact that the Cu$_3$Au
alloy is present almost throughout the entire multilayer as evident
from the SIMS depth profiles (Figure~3b, bottom). Thus it is clear
that the grown multilayer is an interface-dominated structure.

Structural aspects of the multilayer were ascertained from HR-XTEM
and XRD results. The near-crystalline growth of the individual
layers are evident from HR-XTEM results as seen from Figure~4. The
\textit{fcc} nature of the individual layers is evident from the FFT
pattern (Figure~4d) of the marked region of the Au layer. XRD
spectra Figure~3c), on the other hand, exhibited some extra peaks
which were not present in that obtained from samples of our earlier
study~\cite{mycu3au}. Of the different peaks obtained, only two
peaks could be identified with d$_{111}$ of both Au and Cu in their
bulk \textit{fcc} phase. None of the other peaks obtained from the
XRD spectra in the detected 2$\theta$ range could be matched with
the bulk phases of Au, Cu or any of their binary alloys. Most of the
peaks can only be assigned to the \textit{D0$_{23}$} phase of
Cu$_3$Au. In our previous study, we had obtained only one peak
(viz., (111)) of the \textit{D0$_{23}$} phase of Cu$_3$Au and that
was expected to be a weak peak from structure factor calculations.
Here we observe intense peaks of (006), (015) and (114) and a weak
(112) peak in addition to the earlier (111) peak. The diffraction
peaks were assigned using the CELREF Ver. 3.0 package. The lattice
parameters \textit{a} and \textit{c}, obtained for the
\textit{D0$_{23}$} phase are 0.3778 nm and 1.3962 nm respectively.
HR-XTEM image of the interfacial region shows the lattice of the
alloy phase (Figure~4b). The periodicity of the atomic arrangement
becomes clearer from the autocorrelation (Figure~4e) of the marked
region at the interface. The non-\textit{fcc} nature of the lattice
is also evident from the corresponding FFT pattern (Figure~4f). The
dihedral angle estimated from the FFT is found to be 35.23$^{\circ}$
which is quite close to 34.66$^{\circ}$, the angle between the (015)
and (114) planes of the \textit{D0$_{23}$} phase. The calculated
ratio d$_{015}$/d$_{114}$ (= 1.058) for the \textit{D0$_{23}$} phase
also matches with that (= 1.0684) obtained from the FFT results. The
\textit{D0$_{23}$} phase of the Cu$_3$Au alloy is now confirmed. To
our knowledge, this is the first time that such a non-bulk alloy
phase of gold and copper could be grown at ambient temperature in
substantial amounts, although such a phase was predicted earlier
from LDA calculations~\cite{zun}. In addition to the
\textit{D0$_{23}$} phase, Cu$_3$Au in the \textit{L1$_{2}$} phase
with a lattice constant of 0.3778 nm (quite different from its bulk
value of 0.3749 nm) was also found to exhibit its (111) peak. The
analysis of the HR-XTEM image of the Au lattice and the
\textit{D0$_{23}$} phase shows that the (015) planes of the alloy
phase grows parallel to the (220) planes of the Au layer. However,
it is to be noted that due to insufficient TEM data we are unable to
confirm this orientational assignment.

\subsection{Annealing of the novel alloy phase}

We then proceeded to see the thermal behaviour of this novel alloy
phase. The samples were vacuum annealed ($\sim$ 1$\times$10$^{-5}$
mbar using a turbo-molecular pump, Pfeiffer Vacuum) for about two
hours each, in a quartz tube within a Thermolyne tube furnace. The
heating rate was 1$^\circ$C/min. The alloy was annealed at
100$^{\circ}$C, 125$^{\circ}$C and 150$^{\circ}$C. Results
(Figure~5a) show distinct changes in the XRD profile as the
temperature increases. It is evident from the figure that the peaks
corresponding to the \textit{D0$_{23}$} phase tend to become more
intense as the temperature approaches 150$^{\circ}$C. Also, with the
increase of temperature we see the emergence and growth of the (111)
peak of CuAu. Figure~5b presents the results of annealing of the
film quantitatively in a plot of the XRD peak intensities
(normalized with respect to that of Au (111) at room temperature)
versus temperature. The plots show three major trends as the
temperature rises, (a) Au (111) and Cu (111) peaks grow weaker, (b)
CuAu (111) peak grows stronger quite rapidly and (c) strong peaks of
\textit{D0$_{23}$} Cu$_3$Au namely, (114), (015) and (006), as well
as the (111) peak of \textit{L1$_{2}$} Cu$_3$Au grow stronger at a
slow but steady rate. Also the weak peaks, (111) and (112) of
\textit{L1$_{2}$} Cu$_3$Au grow weaker as the temperature rises.
From these trends we can infer that as annealing temperature
approaches 150$^\circ$C: (a) Au and Cu in the multilayer convert to
alloys, an inference supported by (b) the rapid growth of the CuAu
alloy and (c) the steady growth of the metastable
(\textit{D0$_{23}$}) and stable (\textit{L1$_{2}$}) forms of the
Cu$_3$Au alloy. The rapid growth of CuAu is expected as this alloy
corresponds roughly to the atomic ratio of Au and Cu in the
multilayer and would be the probable final state of the multilayer
film upon annealing at higher temperature~\cite{massalski}. What is
interesting is the `stabilization' of the metastable alloy
\textit{D0$_{23}$} at 150$^\circ$C. The lattice parameter
\textit{a}, estimated from our XRD results is 0.3778 nm. This
remains unchanged upto 150$^\circ$C. The other lattice parameter
\textit{c}, varies from 1.3962 nm to 1.3952 nm in the 27$^\circ$C --
150$^\circ$C temperature range. Therefore, the direction of
relaxation of this phase is along its \textit{c} direction, the
direction of tetragonal distortion. It is to be noted that this
stabilization or relaxation is taking place at about 240 degrees
lower than the bulk \textit{L1$_{2}$} phase stabilization
temperature, indicating a shallower minimum in enthalpy as compared
to the \textit{L1$_{2}$} phase, consistent with the results of the
LDA calculation~\cite{zun}. Also such low temperature
`stabilization' of metastable phases can be caused by local
attractive potentials that utilize the small extra movement caused
by annealing to `trap' more atoms thereby forming the alloy in the
metastable phase. The considerable strain at the interfacial region
caused by this local potential is indicated by the shift of the
(111) peak of \textit{L1$_{2}$} Cu$_3$Au from its bulk value. It is
found that the lattice parameter \textit{a}, has a value of 0.3778
nm upto 150$^\circ$C. Upon further annealing, this reduces very
slowly. It is interesting to note in this context that the lattice
parameter of the \textit{L1$_{2}$} phase always matches (even beyond
150$^\circ$C) the lattice parameter \textit{a}, of the
\textit{D0$_{23}$} phase indicating again that the relaxation occurs
only along the \textit{c} axis. The loss in strength of the (111)
and (112) peaks of the \textit{D0$_{23}$} phase with annealing is
again expected as these peaks would be forbidden for a stable and
relaxed \textit{D0$_{23}$} phase. Above 150$^\circ$C, CuAu formation
becomes dominant and neither any new alloy phases nor any
interesting development of the \textit{D0$_{23}$} phase is observed.
It is to be noted that peaks due to oxides of copper are absent
throughout the annealing process.

\subsection{Factors affecting the alloy growth}

In order to have an idea on the formation mechanism of this new
alloy phase, we take a closer look at the growth process. The
deposition wattage used for the growth of the layers serves a
two-fold purpose in this context.

\medskip
\noindent (1) It determines the width of the interfacial region in
the multilayer system.

\medskip
\noindent (2) More importantly, it supplies the amount of energy
required for the formation of the alloy phase.

\medskip
The first point is very important because the alloy is obtained
only at the interfaces of the sequentially deposited Au/Cu system.
We do not get this metastable phase when Au and Cu are
co-sputtered under identical deposition conditions~\cite{mycu3au}.
There we get the bulk \textit{L1$_{2}$} phase of Cu$_3$Au. The
tetragonal phase formed at the multilayer interfaces is obtained
due to the requisite amount of strain (arising out of the
interdiffusion of Au and Cu atoms for the particular interfacial
width) maintained throughout the multilayer. It is to be borne in
mind that the strain will have a gradient across the interface,
which in turn is determined by the stoichiometric gradient of Au
and Cu atoms at the interface. Earlier investigations reveal that
an in-plane strain will produce a strain in the out-of-plane
direction which eventually turns out to be a long period
superlattice (A$_5$BAB) direction along the \textit{c}-axis in our
case~\cite{zun}. Minimization of the interfacial in-plane strain
should bring the in-plane lattice constant \textit{`a'} to a value
in between those of bulk Cu and Au, as is observed from our
annealing experiments.

It may be worth mentioning that the formation of this alloy phase
in the present case is due to interfacial mixing that is
determined by the deposition conditions and interdiffusion does
not probably play a major role. However, with the increase of
temperature interdiffusion has been found to occur across the
interfaces of these multilayers~\cite{madakson}. Accordingly, in
the temperature range of our study, diffusion takes place along
defects and grain boundaries present in the system and
consequently, the alloy formed is seen to persist even up to
150$^\circ$C. Alloying due to interdiffusion has been observed to
dominate for Cu/Au systems after about
250$^\circ$C~\cite{madakson}. However, we have noticed the
formation of a very small amount of CuAu alloy to start from about
100$^\circ$C.

In order to understand the second point in more detail, we have to
look at the magnetron sputtering phenomenon. The hyperthermal
species (sputtered atoms and reflected plasma gas neutral atoms)
formed during the deposition process play a significant role in
the growth process in terms of the crystallinity, microstructure
etc. of the films. Earlier works indicate that metastable phases
can also be formed due to these species~\cite{rabalais1,
rabalais2}. In dc-sputtering deposition systems, the growing film
is mainly bombarded by sputtered target atoms and reflected
neutral plasma gas atoms. These hyperthermal species have their
energies in the range 1 - 1000 eV. The intensity of the
bombardment is determined by the growth conditions: the chamber
pressure, deposition wattage, geometry of the deposition system,
etc. The mean free path ($\lambda$) of the Ar ions depends upon
the working pressure \textit{P} and is given by~\cite{jacob}

\begin{equation}
\lambda_{Ar^+} \approx 1.1/P
\end{equation}

\noindent which is 2.2 cm for our case. In order to have reflected
neutrals, the mean free path of the Ar atoms should be greater
than the cathode dark space whose length \textit{d} (cm) is given
by Child-Langmuir equation~\cite{langmuir}

\begin{equation}
d^2=8.6 \times 10^{-9}~V^{3/2}/J
\end{equation}

\noindent where \textit{V} (in V) is the dc voltage applied to the
cathode and \textit{J} (in A/cm$^2$) is the ion-current density.
Under 25 watts deposition condition d$_{Au}$ = 0.1803 cm and
d$_{Cu}$ = 0.1312 cm. Thus it is evident that the Ar atoms easily
surpass the cathode dark space. Now the reflection coefficient
($\rho$) is given by the relation

\begin{equation}
\rho=1-\frac{m}{M}
\end{equation}

\noindent where \textit{m} and \textit{M} are the masses of the
incident ion (i.e. Ar) and the sputtered atom (i.e. Au or Cu),
respectively. It turns out that the reflection coefficient of Ar
from Au target (\textit{$\rho$$_{Au}$}) is 0.7969 while that for
Cu target (\textit{$\rho$$_{Cu}$}) is 0.37. Thus it is clear that
majority of the reflected neutrals arise when Au sputtering
occurs. According to Kaminsky~\cite{kaminsky} the maximum and
minimum energy of the reflected species are given by the relations

\begin{equation}
E_r^{max} = \left(\frac{M-m}{M+m} \right) E
\end{equation}
\begin{equation}
E_r^{min} = \left(\frac{M-m}{M+m} \right)^2 E
\end{equation}

\noindent where $E$ is the energy of the incident ion. The values
calculated (for 25 watts deposition) using the above equations for
Au and Cu gives 292 eV and 83 eV respectively for
\textit{E$_r$$^{max}$} and 193 eV and 19 eV respectively for
\textit{E$_r$$^{min}$}. The \textit{E$_r$$^{max}$} and
\textit{E$_r$$^{min}$} values for Au are more sensitive to the
change in deposition wattage in comparison to Cu. These reflected
neutrals then tend to thermalize as they travel towards the
growing film. The extent of thermalization is given by the energy
loss of the reflected atom as it passes through the sputtering gas
and can be estimated from the relation~\cite{meyer}

\begin{equation}
E_{final}=(E_0-k_BT_G)exp\left[\frac{PD\sigma}{k_BT_G}ln(E_f/E_i)\right]+k_BT_G
\end{equation}

\noindent where $E_0$ is the energy of the reflected atom as it
leaves the target, $T_G$ is the sputtering gas temperature, $P$ is
the sputtering gas pressure, $D$ is the target-substrate distance,
$\sigma$ is the collision cross-section assuming hard core
interactions and $E_f/E_i$ is the ratio of energies before and
after a collision. Thus it follows from the above equation that
for a particular sputtering gas and sputtered atom combination the
pressure-distance product ($PD$), which is 55 Pa-mm for our case,
determines the extent of thermalization. Therefore, the
thermalized atoms will finally have energies in the region 250 --
350 eV~\cite{somekh} while the mean bombarding energy per arriving
Ar atom may be $\sim$ 20 -- 25 eV~\cite{jacob} or even less. The
last values have been estimated from Ref.~\cite{jacob} where
thermalization was calculated for sputtered C atoms. The reflected
Ar atoms will thermalize even more effectively due to equal mass.
Moreover, it is important to remember that some phenomena such as
transport efficiency, bombardment frequency, electronic
excitation, ionization etc. were not considered in the
calculations. Although our calculations overestimate the real
experimental values, it can still be argued that the reflected
neutrals have sufficient energy necessary to provide the enthalpy
of formation of the new alloy phase at the interfaces. This takes
place most effectively at about 25 watts. We argue that the
deposited energy is low for deposition at 13 watts while it is far
higher at 50 watts deposition condition. This is somewhat similar
to our observations while we anneal the sample. At lower
temperatures there is very little change of the phase. The phase
stabilizes at about 150$^{\circ}$C. Beyond this temperature the
structure breaks down and we tend to get an CuAu bulk alloy.

\section{Conclusion and Outlook}

We have shown that a metastable \textit{D0$_{23}$} phase of
Cu$_3$Au can be grown in substantial amounts by dc-magnetron
sputtering. Interfacial confinement primarily drives this alloy
formation. Additionally, hyperthermal atoms present in the
deposition chamber help in the interface-dominated growth of the
novel alloy phase. Our annealing experiments reveal that the alloy
tends to stabilize as the temperature is increased to about
150$^\circ$C which is 240 degrees below the annealing temperature
of the bulk \textit{L1$_2$} phase of Cu$_3$Au.

Our method of interface-dominated alloy growth can, in principle, be
used to grow metastable alloys to any desired amount thus enabling
one to study the properties of such alloys in greater detail. The
nature of the interface potential and the exact energetics of the
arriving hyperthermal species need to be understood to better
control the growth of the present and other metastable alloy phases.
The presence of such alloys, including the hitherto unobserved
CuAu$_2$ alloy, has been predicted from the same LDA
calculations~\cite{zun} and may be indicated by the presence of new
diffraction peaks that could not be assigned (Figure~3c) in the
present work. Such studies are underway.

\newpage

\begin{table}
\centering \caption{Intensity of \textit{D0$_{23}$} Cu$_3$Au (111)
peak (peak to background ratio) and interfacial widths of films
deposited under different wattages. \label{table:watt}}
\vspace{0.2in}
\begin{tabular}{|c|c|c|} \hline
Deposition wattage & Cu$_3$Au (111) Intensity$^*$ & Interfacial width \\
\small{(watts)} & \it{(from XRD)} & \it{(from SIMS)} \\ \hline
13 & 1.103 & 6.0 nm \\ \hline
25 & 1.203 & 6.8 nm \\ \hline
50 & 1.000 & 8.0 nm \\ \hline
\end{tabular}
\end{table}
$^*$Relative to intensity at 50 watts deposition power.

\vspace*{0.1in}

\newpage

\textbf{Figure Captions} \vspace*{0.25in}

\noindent Figure 1.: (a) Cross-sectional TEM image of the
(Cu/Au)$_{10}$/Si multilayer, (b) SIMS depth profiles showing Au, Cu
and Si signals in a (Au/Cu)$_2$/Si film; Inset: SIMS depth profiles
showing Cu and Cu$_2$Au alloy fragment signals in the same film, (c)
XRD spectrum of the (Au/Cu)$_{10}$/glass multilayer and (d) Crystal
structures of \textit{L1$_{2}$} and \textit{D0$_{23}$} structures.

\vspace*{0.1in}

\noindent Figure 2.: (a) XRD spectra of a Au-capped
(Cu/Au)$_{2}$/glass multilayer deposited at (a) 13 watts, (b) 25
watts and (c) 50 watts. (b) SIMS depth profiles showing Au and Cu
signals of a Au-capped (Cu/Au)$_{2}$/glass multilayer deposited at
(a) 13 watts, (b) 25 watts and (c) 50 watts. The large humps in the
Au signal are due to matrix effect~\cite{myoxy}.

\vspace*{0.1in}

\noindent Figure 3.: (a) Cross-sectional TEM image of the
(Cu/Au)$_{20}$/Si multilayer, (b) Top: SIMS depth profile showing Cu
signal of a Au-capped (Cu/Au)$_{20}$/glass multilayer; Bottom: SIMS
depth profiles showing Cu and Cu$_2$Au alloy fragment signals up to
7 layers of the same multilayer and (c) XRD spectrum of the
Au-capped (Cu/Au)$_{20}$/glass multilayer.

\vspace*{0.1in}

\noindent Figure 4.: (a) HR-XTEM image of a Au layer showing the
well-ordered lattice planes, (b) HR-XTEM image of an interfacial
region of the multilayer, (c) Autocorrelation of the marked region
of (a), (d) FFT of the marked region of (a), (e) Autocorrelation of
the marked region of (b), (f) FFT of the marked region of (b).

\vspace*{0.1in}

\noindent Figure 5.: (a) XRD results of annealing at different
temperatures of the Au-capped (Cu/Au)$_{20}$/glass multilayer, (b)
Intensity of the peaks at different temperatures of the Au-capped
(Cu/Au)$_{20}$/glass multilayer.

\end{document}